\documentstyle[12pt]{article}
\def\lapproxeq{\lower .7ex\hbox{$\;\stackrel{\textstyle <}{\sim}\;$}}
\def\gapproxeq{\lower .7ex\hbox{$\;\stackrel{\textstyle >}{\sim}\;$}}

\begin{document}

\titlepage

\begin{flushright} DTP/93/48 \\ RAL-93-047 \\ August 1993
\end{flushright}

\begin{center}
\vspace*{2cm}
{\large{\bf Two-jet hadroproduction as a measure of the gluon at small $x$}}
\end{center}

\vspace*{.75cm}
\begin{center}
A.D.\ Martin and W.J. Stirling \\
Department of Physics, University of Durham, \\ Durham DH1 3LE, England.  \\
\end{center}

\begin{center}
and
\end{center}

\begin{center}
R.G.\ Roberts \\
Rutherford Appleton Laboratory, \\ Chilton, Didcot OX11 0QX, England. \\
\end{center}

\vspace*{1.5cm}

\begin{abstract}
We investigate the proposal of the CDF collaboration that same-side two-jet
production in $p\bar{p}$ collisions may be used to determine the gluon
distribution at small $x$.
\end{abstract}

\newpage

The gluon is the least well constrained of all the parton distributions of the
proton, although it dominates at small $x$.  There are essentially only two
reliable and precise constraints \cite{MRS1}.  First the measurements of
deep-inelastic lepton-nucleon scattering determine the total fraction of the
proton's momentum that is carried by the gluon.  Secondly the WA70
measurements \cite{WA70} of the prompt-photon reaction, $pp \rightarrow \gamma
X$, determine the gluon in the region $x \approx$ 0.35.

The behaviour of the gluon at small $x$ is particularly important
phenomenologically, but it is also interesting in its own right.  The
resummation of soft gluon emission, as embodied in the Lipatov (or BFKL)
equation \cite{LIP}, implies that $xg \sim x{-\lambda}$ as $x \rightarrow 0$
with $\lambda \approx 0.5$.  Of course as $x$ decreases we will reach a stage
where this growth is suppressed by gluon shadowing effects, and eventually we
enter the confinement region where perturbative QCD ceases to be valid.

The recent preliminary measurements \cite{H1} of the structure function
$F_2(x,Q2)$ for deep-inelastic electron-proton scattering at HERA do show
evidence of a Lipatov-type growth for $x \sim 10{-3}$.  This hints that the
sea quark distribution has the behaviour $x\bar{q} \sim x{-\lambda}$, and
thus, implicitly, also the gluon (if the sea quarks are, as we expect, driven
by $g \rightarrow q\bar{q}$).  Although the above conclusion is plausible, it
is clear that a direct measurement of the gluon at small $x$ is urgently
needed.  Extrapolations of partons to small $x$ give widely differing gluon
distributions.  For the purposes of illustration we take the D${\prime}_0$
and D${\prime}_-$ parton distributions of ref.\ \cite{MRS} which both give
equally acceptable descriptions of fixed target deep-inelastic and related
data.  Although D${\prime}_-$ is favoured by the new preliminary HERA
measurements, recall that the data test $x\bar{q} \sim x{-0.5}$ and not the
gluon.  At $Q$ = 5 GeV the D${\prime}_0$ and D${\prime}_-$ gluons differ by
about a factor of 2 at $x \sim 10{-3}$.  When evolved up in $Q2$ the two
distributions become more similar as can be seen from the results for $Q$ = 15
GeV that are shown in Fig.\ 1.  It is interesting to note that the Lipatov
growth of $xg$ of D${\prime}_-$ must be compensated by a cross-over with the
D${\prime}_0$ gluon so that the total momentum carried by the gluon is
essentially the same for both distributions.  In this note we study the
possibility of using 2-jet production in $p\bar{p}$ collisions to determine
the behaviour of the gluon at small $x$ and, in particular, of distinguishing
between the distributions in Fig.\ 1.

The two-jet cross section may be written to leading order in terms of the sum
of  $i + j \rightarrow k + \ell$ partonic subprocesses
\begin{equation}
\frac{d2\sigma}{dy_1dy_2dp2_T} \; = \; \frac{1}{16\pi s2}
\sum_{i,j,k,\ell} \frac{f_i(x_1,\mu)}{x_1} \, \frac{f_j(x_2,\mu)}{x_2}
\overline{\sum} |{\cal M}(ij \rightarrow k\ell)|2
\end{equation}
where $f_i$ are the parton densities of type $i = g,u,\bar{u},d ...$ evaluated
at momentum scale $\mu$, and $y_1,y_2$ are the laboratory rapidities of the
outgoing partons each of transverse momentum $p_T$.  $\overline{\sum}|{\cal
M}|2$ represents the sub-process matrix elements squared averaged over
initial, and summed over final, parton spins and colours.  For the moment we
assume we can identify the outgoing jets with the outgoing partons.  The
observed jet rapidities can be used to determine the laboratory rapidity of
the two-parton system ($y_{{\rm boost}}$) and the equal and opposite
rapidities ($\pm y*$) of the two jets in the parton-parton c.m.\ frame
\begin{equation}
y_{{\rm boost}} \; = \; {\textstyle \frac{1}{2}} (y_1+y_2) , \hspace*{1cm} y*
\; = \; {\textstyle \frac{1}{2}} (y_1-y_2) .
\end{equation}
The longitudinal momentum fractions of the incoming partons are then given by
\begin{equation}
x_{1,2} \;\; = \;\; \frac{2p_T}{\sqrt{s}}\; {\rm cosh}(y*)\; {\rm exp}(\pm
y_{{\rm boost}}) .
\end{equation}

The CDF collaboration \cite{CDF} have emphasized that their observation of a
pair of same-side jets with large and equal rapidities $y_1 = y_2$ can give a
valuable determination of the gluon density at small $x$.  For example for
same side jets with $y_1 = y_2 = 2.5$ and $p_T = 35$ GeV at $\sqrt{s} = 1.8$
TeV we have
\begin{equation}
x_1 \; = \; 0.47 , \hspace*{1cm} x_2 \; = \; 0.0032 .
\end{equation}
For these $x$ values the jet-pair will originate from $q_{{\rm val}}(x_1)
g(x_2)$ and so an accurate  measurement of same side jet production will be a
valuable determination of the gluon at small $x$, a region in which it is so
far unmeasured. The idea is similar to that proposed \cite{MS} for forward
$Z0$ production, but has the added advantages of a variable dijet mass
and higher statistics.

Before such a method can be employed we must address two problems.  On the
experimental side there are uncertainties arising from normalization, jet
trigger efficiency and energy smearing etc.  On the theoretical side there are
ambiguities associated with the choice of scale $\mu$.  To overcome the
experimental problem the CDF collaboration \cite{CDF} normalise the signal to
the production of a pair of identical jets but with opposite rapidities, $y_1
= -y_2$.  That is they measure the ratio
\begin{equation}
R_1(y,p_T) \; \equiv \; \frac{\sigma_{{\rm SS}}}{\sigma_{{\rm OS}}} \; = \;
\frac{\mbox{No. of same-side jets  (with  $y_1 = y_2 =y$) }}
     {\mbox{No. of opposite-side jets  (with  $y_1 = -y_2 =y$) }}
\end{equation}
{}From (2) we see that $\sigma_{{\rm SS}}$ is built up from parton cross
sections with $y_{{\rm boost}} = y$ and $y* = 0$, whereas $\sigma_{{\rm OS}}$
corresponds to $y_{{\rm boost}} = 0$ and $y* = y$.  Hence the opposite-side
jets originate from partons each with $\bar{x} = x_1 + x_2$ where $x_1$ and
$x_2$ are the momentum fractions of the partons giving the same-side jets.

To relate the measured ratio $R_1(y,p_T)$ to the parton distributions we must
choose the scale $\mu$ in (1).  Since we wish to use $R_1(y,p_T)$ to
distinguish between gluon distributions with $g(x \sim 0.003, \mu \sim p_T)$
which differ by about 30\%, this is clearly an important issue.  To study the
scale dependence we draw on the work of Ellis, Kunszt and Soper \cite{EKS} on
2-jet production at $O(\alpha_s3)$.  The $O(\alpha3_s)$ calculation reduces
the dependence on the choice of scale.  Inter alia, Ellis et al.\ determine the
scale $\mu$ for which the Born or lowest-order $(O(\alpha_s2))$ calculation
approximately reproduces the less scale dependent $O(\alpha3_s)$ result.
They find
\begin{equation}
\mu \; \approx \; \frac{{\rm cosh}(y*)}{{\rm cosh}(0.7y*)} \, \frac{p_T}{2}
\; \equiv \; k(y*) \frac{p_T}{2} ,
\end{equation}
so for same-side jets $\mu = p_T/2$ whereas for opposite side jets $k$
increases from 1 to 2.4 as $y*$ goes from 0 to 3.  In terms of partons we
therefore have
\begin{equation}
R_1(y,p_T) \;\; = \;\; \frac{\sum_{i,j} f_i(x_1,\frac{1}{2}p_T)
f_j(x_2,\frac{1}{2}p_T) \alpha2_s(\frac{1}{2}p_T) \hat{\sigma}_{ij}(0,p_T)}
{\sum_{i,j} f_i(\bar{x},\frac{1}{2}kp_T) f_j(\bar{x},\frac{1}{2}kp_T)
\alpha2_s(\frac{1}{2}kp_T) \hat{\sigma}_{ij}(y,p_T)}
\end{equation}
where we have extracted the $\alpha2_s$ factors from the subprocess cross
sections $\hat{\sigma}(ij \rightarrow 2$ partons).  Here $\bar{x} = x_1 + x_2$
and $k(y)$ is given by (6).

When the $x_i$ in (7) are small, we would expect the cross section
to be dominated by gluon-gluon  scattering. Conversely, when the
$x_i$ are large, valence quark scattering will dominate. More quantitatively,
we recall the \lq single effective subprocess approximation' \cite{SESA}, which
states that the gluon-gluon, quark-gluon and quark-quark subprocess scattering
are approximately in the ratio $1:{\textstyle\frac{4}{9}} :
 {\textstyle\frac{4}{9}}2$. The numerator and denominator in (7) can
therefore be
approximated by $F(x_1) F(x_2) \alpha_s2 \hat{\sigma}_{gg}$, where
$F(x) = g(x) + {\textstyle\frac{4}{9}} \sum_q ( q(x)+\bar q(x) )$.
It is then straightforward to identify the dominant subprocesses at given
$y$ and $p_T$.
In particular,
for large $y$ the observed ratio $R_1(y,p_T)$ directly measures the gluon
distribution $g(x,\mu)$ at small $x$.  To be precise
\begin{equation}
R_1(y,p_T) \;\; \approx \;\; C\; g(x_2, {\textstyle \frac{1}{2}} p_T)
\end{equation}
where $x_2 = 2p_T e{-y}/\sqrt{s}$ and $C(y,p_T)$ depends on parton
distributions at $x$ values where they are reliably known.  This is not quite
true because there is some uncertainty in $C$ from the lack of knowledge of
$g(\bar{x},\frac{1}{2}kp_T)$.

The predictions for $R_1$ obtained using the MRS parton sets D${\prime}_0$
and D${\prime}_-$ are compared with a preliminary sub-sample of CDF data in
Fig.\ 2.  We see that the data definitely favour the D${\prime}_-$ small $x$
behaviour of the gluon in comparison with that for D${\prime}_0$,
although this should be regarded as an illustrative comparison since further
detector corrections to the data have to be made.
It is interesting to note that if we were to take $k = 1$ in the denominator
of (7) then the peak values of D${\prime}_-$ and D${\prime}_0$ in Fig.\ 2
would be reduced to 1.8 and 1.4 respectively.  This large reduction
demonstrates the importance of the choice of scale.  The main problem is that
at large rapidity the $O(\alpha3_s)$ prediction of the opposite-side jet
cross section itself becomes much more scale dependent (see, for example,
Fig.\ 2 of Ellis et al.\ \cite{EKS} and note that for $y = 2$ their variable
$\lambda \equiv  \frac{1}{2}$ sinh$(2y) \simeq 14)$.  This scale dependence of
$\sigma_{{\rm OS}}$ would appear to make it problematic to use the ratio (5)
to definitively measure the gluon to much better than 30\%.  We should also
note that we are applying formula (6) of Ellis et al.\ outside the kinematic
region for which it was established.  Thus for $\sigma_{{\rm SS}}$ the dijet
mass $M_{jj} = 2p_T$ is too small, while for $\sigma_{{\rm OS}}$ for $y
\gapproxeq 2$ the value of  $\lambda$ becomes too large.  However the general
trends are clear.

In an attempt to reduce the scale dependence we introduce an alternative ratio
\begin{displaymath}
R_2(y,p_T) \; \equiv \; \frac{\sigma_{{\rm SS}}(y)}{\sigma_{{\rm SS}}(0)} \; =
\;
\frac{\mbox{No. of same-side jets  (with  $y_1 = y_2 =y$ )} }
     {\mbox{No. of central jets  (with  $y_1 = y_2 =0$) }}
\end{displaymath}
\begin{equation}
= \; \frac{\sum_{i,j} f_i(x_1,\frac{1}{2}p_T) f_j(x_2,\frac{1}{2}p_T)
\hat{\sigma}_{ij}(0,p_T)} {\sum_{i,j} f_i(\hat{x},\frac{1}{2}p_T)
f_j(\hat{x},\frac{1}{2}p_T) \hat{\sigma}_{ij}(0,p_T)}
\end{equation}
where now $\hat{x} = \sqrt{x_1x_2} = 2p_T/\sqrt{s}$.  For $R_2$,
the subprocess cross sections in the numerator and in the denominator
are evaluated at exactly the {\it same} centre-of-mass kinematics,
and therefore we would expect almost all of the scale dependence uncertainty
to cancel. In fact, only a weak
dependence on the factorization scale would remain.
Although this ratio also
partially removes the experimental uncertainties, to reliably measure the
ratio of forward to central 2-jet production presents more of a challenge than
$R_1$ of (5), since the jet pairs at large $y$ are close to the beam.
 At large $y$, a measurement of $R_2$ determines
\begin{equation}
R_2(y,p_T) \;\; \approx \;\; C{\prime}  \;
\frac{g(x_2,\frac{1}{2}p_T)}{g2(\hat{x},\frac{1}{2}p_T)}
\end{equation}
where $C{\prime}(y,p_T)$ is known, $\hat{x} = 2p_T/\sqrt{s}$ (= 0.039 for
$p_T$ = 35 GeV and $\sqrt{s}$ = 1.8 TeV) and $x_2 = \hat{x} e{-y}$ (= 0.0056
for $y$ = 2 for example).  In Fig.\ 3 we compare the predictions of $R_2$
obtained from parton sets D${\prime}_-$ and D${\prime}_0$ \cite{MRS}.  The
D${\prime}_-$ prediction for $R_2$ is more than 30\% larger than that for
D${\prime}_0$ for $y \approx 2$ due partly to the numerator of (10), but
mainly due to the quadratic dependence in the denominator.  Both factors
increase $R_2$(D${\prime}_-)/R_2($D${\prime}_0)$ as can be seen from the
different behaviours of $xg$ for D${\prime}_0$ and D${\prime}_-$ shown in
Fig.\ 1, with a cross-over at $x \sim 0.009$ required to maintain the total
momentum carried by the gluon.

As compared to $R_1(y,p_T)$, the ratio $R_2(y,p_T)$ has the advantages of (i)
much
less scale dependence and (ii) large differences due to the different small
$x$ behaviours of the gluon at more moderate values of $y$, that is $1.5
\lapproxeq y \lapproxeq 2$.  We conclude that measurements of same-side 2-jet
production at the FNAL $p\bar{p}$ collider offer the possibility of
determining the small $x$ behaviour of the gluon, but that the uncertainties
in the choice of scale need to be carefully considered.  In general the scale
dependence can be reduced by considering jets of higher $p_T$.  This would
offer a valuable constraint on the gluon at somewhat higher values of $x$.
The preliminary sub-sample of CDF 2-jet data appear to favour MRS parton set
D${\prime}_-$, as compared to D${\prime}_0$; a result also found from
measurements of $F_2(x,Q2)$ at HERA \cite{H1}.
When final CDF data are available, a precise determination of the
behaviour of the gluon at small $x$ should be possible.

\vspace*{1cm}
\noindent {\bf Acknowledgements}
\vspace*{.5cm}

We are grateful to Eve Kovacs and Robert Plunkett for drawing this problem to
our attention and for information concerning their
data, and to the CDF collaboration for allowing us to illustrate our
discussion using a preliminary sub-sample of their data.

\vskip 1cm
%\newpage
\noindent{\Large\bf Figure Captions}

\begin{itemize}
\item[Fig.\ 1] The gluon $xg(x,Q$ = 15 GeV) corresponding to the
D${\prime}_-$ and D${\prime}_0$ set of partons of ref.\ \cite{MRS}.

\item[Fig.\ 2] The curves show the same-side/opposite-side dijet ratio
$R_1$ predicted from (7) using the D${\prime}_-$ and D${\prime}_0$ partons
of ref.\cite{MRS} for $27 < p_T < 60$ GeV.
The data points are the preliminary CDF measurements \cite{CDF}
of the ratio for jets  with $27 < E_T < 60$ GeV.
The measured $E_T$ values have not been corrected for CDF detector effects
and therefore do not correspond directly to the true jet transverse energies.

\item[Fig.\ 3] The ratio  $R_2$ of (9) calculated using the D${\prime}_-$ and
D${\prime}_0$ partons of ref.\ \cite{MRS} for $27 < p_T < 60$ GeV.

\end{itemize}

\end{document}